\documentclass[12pt]{article}
\usepackage{amsmath,amssymb}

\begin{document}

\title{Actions for spinor fields in arbitrary dimensions}

\author{M. A. De Andrade and I. V. Vancea}

\date{}

\maketitle

We are going to give a systematic presentation of spinors in various spacetime dimensions which is 
a prerequisite for understanding the supersymmetry. (For general reviews see [1,2,3].)

To set up the conventions, we will consider in what follows the Dirac matrices that satisfy the following 
Clifford algebra
\begin{equation*}
\{\gamma ^{\mu },\gamma ^{\nu }\}=2\eta ^{\mu \nu }{\leavevmode\hbox{\small1\kern-3.8pt\normalsize1}}\;,\;\;\eta^{\mu\nu}=\text{diag}(+\cdots+,-\cdots-)\;, 
\tag{1}
\end{equation*}
the sign $+$, $-$ appear $t$, $s$ times which are the numbers of timelike, spacelike directions, respectively.
The Hermitian, complex and transpose conjugation operations are defined as usual [4,5] and are given by 
the relations
\begin{equation*}
\gamma ^{\mu \dagger } =-(-1)^{t}A\gamma ^{\mu }A^{-1}\;,\;
\gamma ^{\mu \ast } =\eta B\gamma ^{\mu }B^{-1}\;,\;
\gamma ^{\mu T} =-\eta (-1)^{t}C\gamma ^{\mu }C^{-1}\;,  
\tag{2}
\end{equation*}
where $\eta=\pm 1$ corresponds to independent choices of charge conjugation matrix. $A$, $B$ and $C$ 
are unitary matrices and the following relations hold among them 
\begin{equation*}
A=\gamma ^{1}...\gamma ^{t} \;,\; B\,^{T}=\varepsilon B \;,\;  C\equiv B\,^{T}\,A \;.
\tag{3} 
\end{equation*}
Here, $1,\ldots,t$ are all timelike directions and 
\begin{equation*}
\varepsilon =\cos \frac{\pi }{4}\left( s-t\right) -\eta \sin \frac{\pi }{4} 
\left( s-t\right)\;. 
\tag{4}
\end{equation*}
Note that the explicit form of matrices $B$ and $C$ depend on the specific representation of the 
Dirac matrices. Also, we assume that $s-t$ is always an even number. 

The most general action for free spinors is given by
\begin{equation*}
S=\int d^{D}x\left[\alpha \overline{\Psi }{\rlap{\hbox{$\mskip 1 mu /$}}\partial}
\Psi + \beta m\overline{\Psi }\Psi + \delta m_5\overline{\Psi }
\gamma_{D+1}\Psi\right]\;,    
\tag{5}
\end{equation*}
where $\alpha$, $\beta$ and $\delta$ are determined by the reality of the action and $\overline{\Psi }=\Psi^\dagger A$.
The first two terms in Eq.(5) reproduce the usual Dirac equation. The last term is a 
pseudo-scalar which can be added to the known action whenever the dimension of spacetime is even to make it 
invariant under the chiral transformations. The mass shell condition that defines the Dirac spinors in any 
spacetime dimensions is 
\begin{equation*}
\alpha ^{2}p^{2}=-\beta ^{2}m^{2}+\delta ^{2}{m_{5}}^{2}\;. 
\tag{6}
\end{equation*}
From the Hermiticity of the Lagrangian one obtains some equations that should be satisfied by the unknown 
coefficients in Eq.(5). They are given by the following relations
\begin{equation*}
\alpha ^{\ast }\left( -1\right) ^{\frac{t\left( t+1\right) }{2}}=\alpha
,\,\,\ \ \,\beta ^{\ast }\left( -1\right) ^{\frac{t\left( t-1\right) }{2}
}=\beta ,\,\,\ \ \,\delta ^{\ast }\left( -1\right) ^{\frac{t\left(
t+1\right) }{2}}=\delta \;. 
\tag{7}
\end{equation*}
From the above equations one can write down the value of $\alpha$, $\beta$ and $\gamma$ for arbitrary number of 
timelike directions 

\begin{centering}

\bigskip

\begin{tabular}{|c|c|c|c|c|}
\hline
$t$ & $0~\mbox{mod}~4$ & $1~\mbox{mod}~4$ & $2~\mbox{mod}~4$ & $3~\mbox{mod}~4$ \\
\hline
$\alpha $ & $\ \ \ 1$ & $\ \ \ i$ & $\ \ \ i$ & $\ \ \ 1$ \\ \hline
$\beta $ & $\ \ \ 1$ & $\ \ \ 1$ & $\ \ \ i$ & $\ \ \ i$ \\ \hline
$\delta $ & $\ \ \ 1$ & $\ \ \ i$ & $\ \ \ i$ & $\ \ \ 1$ \\ \hline
$p^2 $ & $-m^2+{m_5}^2$ & $m^2+{m_5}^2$ & $-m^2+{m_5}^2$ & $m^2+{m_5}^2$ \\ \hline
\end{tabular}

\bigskip

{\bf Table 1}

\end{centering}

The $\cal C$-symmetry of the action (charge-conjugation) imposes restrictions on the 
possible values of the parameter $\eta$. Thus, assuming that 
\begin{equation*}
{\cal L}\left( \Psi ^{C}\right) =\left[ {\cal L}\left( \Psi \right) \right] ^{\ast } \;, 
\tag{8}
\end{equation*}
where the charge conjugations of the fields is defined by 
\begin{equation*}
\Psi ^{C}\equiv {\cal C}\Psi {\cal C}^{-1}=B^{-1}\Psi^{\ast } 
\tag{9}
\end{equation*}
it follows from kinetic, mass and chiral mass term 
\begin{equation*}
\eta ^{t+1}\left( -1\right) ^{\frac{t\left( t+1\right) }{2}}=-1 ~,~
\eta ^{t}\left( -1\right) ^{\frac{t\left( t-1\right) }{2}}=-1 ~,~
\eta ^{t}\left( -1\right) ^{\frac{t\left( t-1\right) }{2}}\left( -1\right)^
{\frac{ D }{2}}=-1 ~,~ 
\tag{10}
\end{equation*}
respectively. Let us analyze the existence of the action for different types of charged spinors in even 
spacetime dimensions. If $t=2k$, $k\in\mathbb{Z}$ then one can use Eq.(10) to write down the following 
relations 
\begin{equation*}
\eta = (-1)^{t/2\;+1} ~,~ (-1)^{t/2}=-1 ~,~ (-1)^{s/2}=-1 \;.
\tag{11}
\end{equation*}
We see that for $t=0\text{ mod }4$ the value of $\eta=-1$, while for $t=2\text{ mod }4$ we obtain $\eta=1$. 
The invariance of the mass term ($m$) is allowed only for $t=2\text{ mod }4$. The chiral mass term ($m_5$) 
is invariant only for $s=2\text{ mod }4$. For $t=2k+1$, the corresponding relations are  
\begin{equation*}
(-1)^{(t-1)/2} = 1 ~,~ \eta = (-1)^{(t+1)/2} ~,~ \eta = (-1)^{(s-1)/2}\;.
\tag{12}
\end{equation*}
We summarize these results in the following table

\begin{centering}

\bigskip

\begin{tabular}{|c|c|c|c|c|c|}
\hline
$t $ & $D=2 $ & $D=4 $ &  $ D=6 $ & $D=8 $ & $D=10 $ \\ \hline
{\small 0} & $
\begin{array}{c}
-1,\partial,m_{5},M \\
1,\,m_{5},{\rlap{\hbox{$\mskip 3 mu /$}}M}
\end{array}
$ & $
\begin{array}{c}
-1,\partial,{\rlap{\hbox{$\mskip 3 mu /$}}M}^{\ast } \\
1,\,{\rlap{\hbox{$\mskip 3 mu /$}}M}^{\ast }
\end{array}
$ & $
\begin{array}{c}
-1,\partial,m_{5},{\rlap{\hbox{$\mskip 3 mu /$}}M} \\
1,\,m_{5},M
\end{array}
$ & $
\begin{array}{c}
-1,\partial,M^{\ast } \\
1,\,M^{\ast }
\end{array}
$ & $
\begin{array}{c}
-1,\partial,m_{5},M \\
1,\,m_{5},{\rlap{\hbox{$\mskip 3 mu /$}}M}
\end{array}
$ \\ \hline
{\small 1} & $
\begin{array}{c}
-1,\partial,m,M^{\ast } \\
1,\partial,m_{5},M^{\ast }
\end{array}
$ & $
\begin{array}{c}
-1,\partial,m,m_{5},M \\
1,\partial,{\rlap{\hbox{$\mskip 3 mu /$}}M}
\end{array}
$ & $
\begin{array}{c}
-1,\partial,m,{\rlap{\hbox{$\mskip 3 mu /$}}M}^{\ast } \\
1,\partial,m_{5},{\rlap{\hbox{$\mskip 3 mu /$}}M}^{\ast }
\end{array}
$ & $
\begin{array}{c}
-1,\partial,m,m_{5},{\rlap{\hbox{$\mskip 3 mu /$}}M} \\
1,\partial,M
\end{array}
$ & $
\begin{array}{c}
-1,\partial,m,M^{\ast } \\
1,\partial,m_{5},M^{\ast }
\end{array}
$ \\ \hline
{\small 2} & $
\begin{array}{c}
-1,\,m^{-},{\rlap{\hbox{$\mskip 3 mu /$}}M} \\
1,\partial,m^{-},M
\end{array}
$ & $
\begin{array}{c}
-1,\,m^{-},m_{5},M^{\ast } \\
1,\partial,m^{-},m_{5},M^{\ast }
\end{array}
$ & $
\begin{array}{c}
-1,\,m^{-},M \\
1,\partial,m^{-},{\rlap{\hbox{$\mskip 3 mu /$}}M}
\end{array}
$ & $
\begin{array}{c}
-1,\,m^{-},m_{5},{\rlap{\hbox{$\mskip 3 mu /$}}M}^{\ast } \\
1,\partial,m^{-},m_{5},{\rlap{\hbox{$\mskip 3 mu /$}}M}^{\ast }
\end{array}
$ & $
\begin{array}{c}
-1,\,m^{-},{\rlap{\hbox{$\mskip 3 mu /$}}M} \\
1,\partial,m^{-},M
\end{array}
$ \\ \hline
{\small 3} &  & $
\begin{array}{c}
-1,\,{\rlap{\hbox{$\mskip 3 mu /$}}M} \\
1,\,m,m_{5},M
\end{array}
$ & $
\begin{array}{c}
-1,\,m_{5},M^{\ast } \\
1,\,m,M^{\ast }
\end{array}
$ & $
\begin{array}{c}
-1,\,M \\
1,\,m,m_{5},{\rlap{\hbox{$\mskip 3 mu /$}}M}
\end{array}
$ & $
\begin{array}{c}
-1,\,m_{5},{\rlap{\hbox{$\mskip 3 mu /$}}M}^{\ast } \\
1,\,m,{\rlap{\hbox{$\mskip 3 mu /$}}M}^{\ast }
\end{array}
$ \\ \hline
{\small 4} &  & $
\begin{array}{c}
-1,\partial,{\rlap{\hbox{$\mskip 3 mu /$}}M}^{\ast } \\
1, {\rlap{\hbox{$\mskip 3 mu /$}}M}^{\ast }\
\end{array}
$ & $
\begin{array}{c}
-1,\partial,m_{5},{\rlap{\hbox{$\mskip 3 mu /$}}M} \\
1,\,m_{5},M
\end{array}
$ & $
\begin{array}{c}
-1,\partial,M^{\ast } \\
1,\,M^{\ast }
\end{array}
$ & $
\begin{array}{c}
-1,\partial,m_{5},M \\
1,\,m_{5},{\rlap{\hbox{$\mskip 3 mu /$}}M}
\end{array}
$ \\ \hline
{\small 5} &  &  & $
\begin{array}{c}
-1,\partial,m,{\rlap{\hbox{$\mskip 3 mu /$}}M}^{\ast } \\
1,\partial,m_{5},{\rlap{\hbox{$\mskip 3 mu /$}}M}^{\ast }
\end{array}
$ & $
\begin{array}{c}
-1,\partial,m,m_{5},{\rlap{\hbox{$\mskip 3 mu /$}}M} \\
1,\partial,M
\end{array}
$ & $
\begin{array}{c}
-1,\partial,m,M^{\ast } \\
1,\partial,m_{5},M^{\ast }
\end{array}
$ \\ \hline
{\small 6} &  &  & $
\begin{array}{c}
-1,\,m^{-},M \\
1,\partial,m^{-},{\rlap{\hbox{$\mskip 3 mu /$}}M}
\end{array}
$ & $
\begin{array}{c}
-1,\,m^{-},m_{5},{\rlap{\hbox{$\mskip 3 mu /$}}M}^{\ast } \\
1,\partial,m^{-},m_{5},{\rlap{\hbox{$\mskip 3 mu /$}}M}^{\ast }
\end{array}
$ & $
\begin{array}{c}
-1,m^{-},{\rlap{\hbox{$\mskip 3 mu /$}}M} \\
1,\partial,m^{-},M
\end{array}
$ \\ \hline
{\small 7} &  &  &  & $
\begin{array}{c}
-1,\,M \\
1,\,m,m_{5},{\rlap{\hbox{$\mskip 3 mu /$}}M}
\end{array}
$ & $
\begin{array}{c}
-1,\,m_{5},{\rlap{\hbox{$\mskip 3 mu /$}}M}^{\ast } \\
1,\,m,{\rlap{\hbox{$\mskip 3 mu /$}}M}^{\ast }
\end{array}
$ \\ \hline
{\small 8} &  &  &  & $
\begin{array}{c}
-1,\partial,M^{\ast } \\
1,\,M^{\ast }
\end{array}
$ & $
\begin{array}{c}
-1,\partial,m_{5},M \\
1,\,m_{5},{\rlap{\hbox{$\mskip 3 mu /$}}M}
\end{array}
$ \\ \hline
{\small 9} &  &  &  &  & $
\begin{array}{c}
-1,\partial,m,M^{\ast } \\
1,\partial,m_{5},M^{\ast }
\end{array}
$ \\ \hline
{\small 10} &  &  &  &  & $
\begin{array}{c}
-1,\,m^{-},{\rlap{\hbox{$\mskip 3 mu /$}}M} \\
1,\partial,m^{-},M
\end{array}
$ \\ \hline
\end{tabular}

\bigskip

{\bf Table 2}

\end{centering}

From this table we can read the following information. In each slot, $-1$ and $+1$ are the
values of $\eta$. At the right hand side of them there are some symbols  of which
meanings are given in the sequel. The presence of $\partial$, $m$ and $m_5$,  
denote the existence of kinetic, mass, and chiral mass, respectively,  all satisfying the 
$\cal C$-symmetry; the absence of one of them indicates that the corresponding term  
is not $\cal C$-symmetric. In this case, that term vanishes as a consequence of the
Majorana constraint. The $(-)$ sign labeling $m$  appears when $-m^2$ contributes to
$p^2$ in the mass-shell constraint as can be read off Table~1. The symbols $M$ and 
${\rlap{\hbox{$\mskip 3 mu /$}}M}$ denote  $\varepsilon =1 $ (Majorana spinors) and $\varepsilon=-1$ 
($SU(2)$-Majorana spinors [4]), respectively. The symbol ($*$) always appears 
when $s-t$ = $0\text{ mod }4$. In this case the left (right) charge conjugate spinors 
and left (right) spinors transform in the same way under the Lorentz transformations. 
For example, $M^*$ denotes that the Majorana-Weyl spinors can be present.    

Let us focus now on odd spacetime dimensions. From the first two relations in Eq.(2) applied to 
$\gamma^{D+1}$, one obtains the value of the parameter $\eta$ as a function of $s-t$
\begin{equation*}
\eta = {(-1)}^{\frac{s-t}{2}+1}    \;.  
\tag{13}
\end{equation*}
Similar reasoning as in the even case leads us to following table 

\bigskip

\begin{centering}

\begin{tabular}{|c|c|c|c|c|c|}
\hline
$t $ & $D=3 $ & $D=5 $ &  $ D=7 $ & $D=9 $ & $D=11 $ \\ \hline
{\small 0} & $
\begin{array}{c}
1,\,{\rlap{\hbox{$\mskip 3 mu /$}}M}
\end{array}
$ & $
\begin{array}{c}
-1,\partial,{\rlap{\hbox{$\mskip 3 mu /$}}M}
\end{array}
$ & $
\begin{array}{c}
1,\,M
\end{array}
$ & $
\begin{array}{c}
-1,\partial,M
\end{array}
$ & $
\begin{array}{c}
1,\,{\rlap{\hbox{$\mskip 3 mu /$}}M}
\end{array}
$ \\ \hline
{\small 1} & $
\begin{array}{c}
-1,\partial,m,M
\end{array}
$ & $
\begin{array}{c}
1,\partial,{\rlap{\hbox{$\mskip 3 mu /$}}M}
\end{array}
$ & $
\begin{array}{c}
-1,\partial,m,{\rlap{\hbox{$\mskip 3 mu /$}}M}
\end{array}
$ & $
\begin{array}{c}
1,\partial,M
\end{array}
$ & $
\begin{array}{c}
-1,\partial,m,M
\end{array}
$ \\ \hline
{\small 2} & $
\begin{array}{c}
1,\partial,m^{-},M
\end{array}
$ & $
\begin{array}{c}
-1,\,m^{-},M
\end{array}
$ & $
\begin{array}{c}
1,\partial,m^{-},{\rlap{\hbox{$\mskip 3 mu /$}}M}
\end{array}
$ & $
\begin{array}{c}
-1,\,m^{-},{\rlap{\hbox{$\mskip 3 mu /$}}M}
\end{array}
$ & $
\begin{array}{c}
1,\partial,m^{-},M
\end{array}
$ \\ \hline
{\small 3} & $
\begin{array}{c}
-1,{\rlap{\hbox{$\mskip 3 mu /$}}M}
\end{array}
$ & $
\begin{array}{c}
1,\,m,M
\end{array}
$ & $
\begin{array}{c}
-1,\,M
\end{array}
$ & $
\begin{array}{c}
1,\,m,{\rlap{\hbox{$\mskip 3 mu /$}}M}
\end{array}
$ & $
\begin{array}{c}
-1,\,{\rlap{\hbox{$\mskip 3 mu /$}}M}
\end{array}
$ \\ \hline
{\small 4} &  & $
\begin{array}{c}
-1,\partial,{\rlap{\hbox{$\mskip 3 mu /$}}M}
\end{array}
$ & $
\begin{array}{c}
1,\,M
\end{array}
$ & $
\begin{array}{c}
-1,\partial,M
\end{array}
$ & $
\begin{array}{c}
1,\,{\rlap{\hbox{$\mskip 3 mu /$}}M}
\end{array}
$ \\ \hline
{\small 5} &  & $
\begin{array}{c}
1,\partial,{\rlap{\hbox{$\mskip 3 mu /$}}M}
\end{array}
$ & $
\begin{array}{c}
-1,\partial,m,{\rlap{\hbox{$\mskip 3 mu /$}}M}
\end{array}
$ & $
\begin{array}{c}
1,\partial,M
\end{array}
$ & $
\begin{array}{c}
-1,\partial,m,M
\end{array}
$ \\ \hline
{\small 6} &  &  & $
\begin{array}{c}
1,\partial,m^{-},{\rlap{\hbox{$\mskip 3 mu /$}}M}
\end{array}
$ & $
\begin{array}{c}
-1,\,m^{-},{\rlap{\hbox{$\mskip 3 mu /$}}M}
\end{array}
$ & $
\begin{array}{c}
1,\partial,m^{-},M
\end{array}
$ \\ \hline
{\small 7} &  &  & $
\begin{array}{c}
-1,M
\end{array}
$ & $
\begin{array}{c}
1,\,m,{\rlap{\hbox{$\mskip 3 mu /$}}M}
\end{array}
$ & $
\begin{array}{c}
-1,\,{\rlap{\hbox{$\mskip 3 mu /$}}M}
\end{array}
$ \\ \hline
{\small 8} &  &  &  & $
\begin{array}{c}
-1,\partial,M
\end{array}
$ & $
\begin{array}{c}
1,\,{\rlap{\hbox{$\mskip 3 mu /$}}M}
\end{array}
$ \\ \hline
{\small 9} &  &  &  & $
\begin{array}{c}
1,\partial,M
\end{array}
$ & $
\begin{array}{c}
-1,\partial,m,M
\end{array}
$ \\ \hline
{\small 10} &  &  &  &  & $
\begin{array}{c}
1,\partial,m^{-},M
\end{array}
$ \\ \hline
{\small 11} &  &  &  &  & $
\begin{array}{c}
-1,{\rlap{\hbox{$\mskip 3 mu /$}}M}
\end{array}
$ \\ \hline

\end{tabular}

\bigskip

{\bf Table 3}

\bigskip

\end{centering}

Here, $D=t+s+1$. The symbols in Table 3 have the same meaning as the ones in
the preceding table. However we call the attention to the following details; 
$m_5$ does not appear anymore, because only odd spacetime dimensions 
are shown in Table 3. 
Also, the symbol ($*$), does not appear, since this is related with the Majorana-Weyl 
constraint that does not have any meaning in the case considered here. 
In the Tables 2 and 3
the same symbols repeats for any $t$ and $D\text{ mod }8$ and for any 
$D$ and $t\text{ mod }4$. Therefore, immediately we can infer what happens in 
the spacetime with a number of directions greater than eleven. 

For completeness we give the basic property of the charge conjugation matrix for
arbitrary spacetime dimensions. This is shown in the following formula which 
results from the Eq.(3) 
\begin{equation*}
{\gamma^\mu}^T=\zeta C \gamma^\mu C^{-1}\;,\;C^T=\xi C
\tag{14}
\end{equation*}
where $\zeta$ and $\xi$ can be read off the next table

\begin{centering}

\bigskip

\begin{tabular}{|c|c|c|c|c|}
\hline
$t$ & $0\mbox{mod}4$ & $1\mbox{mod}4$ & $2\mbox{mod}4$ & $3\mbox{mod}4$ \\
\hline
$\zeta $ & $-\eta$ & $\eta$ & $-\eta$ & $\eta$ \\ \hline
$\xi $ & $\varepsilon$ & $\varepsilon\eta$ & $-\varepsilon$ & $-\varepsilon\eta$ \\ \hline
\end{tabular}

\bigskip

{\bf Table 4}
 
\end{centering}

In conclusion, we have given the most general action for free spinor fields in arbitrary spacetime dimensions.
The interaction term 
\begin{equation*}
{\cal L}_{{\rm int}}=i\alpha g\overline{\Psi }\gamma^{\mu}A_{\mu}\Psi \;,  
\tag{15}
\end{equation*}
where $A_{\mu}$ is the gauge potential, does not introduce any new constraint 
on $\alpha$, $\beta$ and $\delta$ as can be easily checked up. 

Supersymmetry in higher dimensions plays a crucial role in string theory based models on high energy physics.
This is the basic justification for most of the supersymmetric field theories and supergravities in 
dimensions other than four.

\subsection*{Bibliography}

\begin{description}
\item{[1] M. A. De Andrade and F. Toppan, {\it Mod. Phys. Lett.} 
{\bf A26} (1999) 1797 - 1814, hep-th/9904134}

\item{[2] A. Van Proeyen, Lectures given at the 
Spring School Q.F.T., Supersymmetry
and Superstrings, Apr 1998, Calimanesti, Romania. hep-th/9910003}

\item{[3] M. A. De Andrade, ``Dirac Spinors in Arbitrary Spacetime'',
http://www.cbpf.br/~dcp/papers/, in portuguese}

\item{[4] T. Kugo and P. Towns
end, {\it Nucl. Phys.} {\bf B221} (1983) 357 - 380}

\item{[5] C. Wetterich, {\it Nucl. Phys.} {\bf B211} (1983) 177 - 188 }
\end{description}

\end{document}